\documentclass[twocolumn,showpacs,prl,a4paper]{revtex4-1}
\usepackage[T1]{fontenc}
\usepackage[latin9]{inputenc}
\usepackage{bm}
\usepackage{amsmath}
\usepackage{graphicx}
\usepackage{esint}
\usepackage[normalem]{ulem}
\usepackage[unicode=true,pdfusetitle,
 bookmarks=true,bookmarksnumbered=false,bookmarksopen=false,
 breaklinks=false,pdfborder={0 0 1},backref=false,colorlinks=false]{hyperref}
\usepackage{hyperref}

\newcommand{\phdag}{\phantom{\dag}}

  \date{May 25, 2012}

\begin{document}

  \author{Markus Greger}

  \author{Marcus Kollar}

  \author{Dieter Vollhardt}

  \affiliation{Theoretical Physics III, Center for Electronic
    Correlations and Magnetism, Institute of Physics, University of
    Augsburg, 86135 Augsburg, Germany}

  \begin{abstract}
    We calculate the spectra and spin susceptibilities of a Hubbard
    model with two bands having different bandwidths but the same
    on-site interaction, with parameters close to the
    orbital-selective Mott transition, using dynamical mean-field
    theory.  If the Hund's rule coupling is sufficiently
    strong, one common energy scale emerges which characterizes both
    the location of kinks in the self-energy and extrema of the
    diagonal spin susceptibilities. A physical explanation of this
    energy scale is derived from a Kondo-type model. We infer
    that for multi-band systems local spin dynamics rather than
    spectral functions determine the location of kinks in the
    effective band structure.
  \end{abstract}

  \title{Emergence of a common energy scale close to the
    orbital-selective Mott transition}

  \pacs{71.27.+a, 71.30.+h}

  \maketitle

  The interactions in correlated metals lead to the emergence
  of characteristic energy scales. Close to the Fermi
  energy Landau Fermi-liquid theory~\cite{pinesnozieres} applies and
  the effective electronic dispersion $E_{\bm{k}}$ is renormalized, but remains linear as
  in the noninteracting case. The linear dependence terminates at an excitation energy which cannot be calculated  within Fermi-liquid theory itself.  With increasing strength of the
  electron-electron interaction this Fermi-liquid coherence scale
  decreases and ultimately vanishes at the Mott transition from a
  metal to an insulator. At the same time charge excitations are
  shifted to higher energies of the order of the interaction energy
  and are thus suppressed.  For the single-band Hubbard model with
  on-site Coulomb interaction $U$ it was shown~\cite{byczuk2006kinks}
  in dynamical mean-field theory
  (DMFT)~\cite{PhysRevLett.62.324, *RevModPhys.68.13, *dvgk_phystoday} that the
  Fermi-liquid regime terminates at an energy scale $\omega_{\Sigma}$
  at which the real part of the self-energy,
  $\text{Re}[\Sigma(\omega)]$, and hence the effective dispersion
  $E_{\bm{k}}$, has a rather sudden change in slope~\footnote{An
    effective dispersion $E_{\bm{k}}$ can always be defined by the
    maxima of the momentum-resolved spectral function
    $A_{\bm{k}}(\omega)$, which is essentially measured in
    angular-resolved photoemission spectroscopy (ARPES), even when the
    spectrum is rather incoherent and no sharp quasiparticle
    excitations occur.}.  This `kink' does not require any coupling to
  external bosonic degrees of freedom but is due to the correlated
  behavior of interacting electrons.  For the single-band model the
  Fermi-liquid scale $\omega_{\Sigma}$ can be derived from the
  low-energy properties of the local spectral
  function~\cite{byczuk2006kinks}.  Moreover, it was
demonstrated~\cite{PhysRevLett.102.076406,grete2011kinks} that the energy
  scale $\omega_{\Sigma}$ is linked to the characteristic energy scale
  $\omega_{\text{sp}}$ of spin fluctuations. Kinks in the electronic
  dispersion were studied theoretically in a variety of
  contexts~\cite{Eschrig_PhysRevLett.85.3261,Manske_PhysRevLett.87.177005,Kakehashi_JPSJ.74.2397,Nekrasov_PhysRevB.73.155112,Held_PhysRevLett.102.076402,Claessen_PhysRevLett.102.187204,lanio3_PhysRevB.85.125137,Bauer_PhysRevB.82.184535}.

  In this Letter we explore the origin and characteristic energy scale
  of kinks in the effective electronic dispersion in a more general
  context. Employing the DMFT we study a two-band Hubbard model with
  two different bandwidths, the same on-site repulsion $U$ for both bands,
  and an interorbital repulsion $U_1$ and ferromagnetic Hund's rule
  spin exchange $J$ between the bands.  Thereby it is possible to capture
  orbital effects in correlated materials that do
  not exist in single-band models. Indeed, different kinks in
  the dispersion depending on the orbital character are observed for
  Sr$_2$RuO$_4$ both experimentally and
  theoretically~\cite{PhysRevB.72.104514,Mravlje_PhysRevLett.106.096401}.

  We study the model Hamiltonian
   \begin{align}
     H
     &=
     \sum_{ijm\sigma}
     t_{ij,m}
     d_{im\sigma}^{\dag}d_{jm\sigma}^{\phdag}
     +
     H_{\text{int}}
     \,,\label{eq:LattHam}
     \\
     H_{\text{int}}
     &=
     U\sum_{im}n_{im\uparrow}n_{im\downarrow}
     +
     \sum_{i\sigma\sigma'}
     (U_{1}-\delta_{\sigma\sigma'}J)
     \,
     n_{i1\sigma}n_{i2\sigma'}
     \nonumber\\&~~~
     +\frac{J}{2}
     \sum_{im\sigma}
     d_{im\sigma}^{\dag}
     (
     d_{i\bar{m}\bar{\sigma}}^{\dag}d_{im\bar{\sigma}}^{\phdag}
     +
     d_{im\bar{\sigma}}^{\dag}d_{i\bar{m}\bar{\sigma}}^{\phdag}
     )
     d_{i\bar{m}\sigma}^{\phdag}
     \,,\nonumber
  \end{align}
  with spin index $\sigma$ $=$ $\uparrow,\downarrow$ and orbital index
  $m$ $=$ $1,2$. Here a bar over an index denotes the opposite spin or orbital. The two bands do not hybridize but are
  coupled by the interorbital interactions $U_1$ and $J$. We
  consider $U$, $U_{1}$ and $J$ as independent parameters which can
  take arbitrary values, but $U_{1}$ $=$
  $U-2J$~\cite{U_minus_2J} for $d$~electrons.  As in the single-band case the
  correlation strengths of the two bands may be roughly parametrized
  by the ratios $U/W_1$ and $U/W_2$, which are assumed to be unequal.
  Due to this difference in the relative interaction strengths an
  orbital-selective Mott transition (OSMT) occurs upon increase of
  $U$~\cite{%
    anisimov2002orbital,%
    springerlink:10.1140/epjb/e2005-00117-4,
    Liebsch_PhysRevLett.95.116402, *Liebsch_PhysRevLett.99.236404,%
    Bluemer_PhysRevB.72.081103,%
    Jakobi_PhysRevB.80.115109,%
    Arita_PhysRevB.72.201102,
    Biermann_PhysRevLett.95.206401,
    Medici_PhysRevB.72.205124,
    Medici_PhysRevB.83.205112,
    Kawakami_PhysRevLett.92.216402, *Kawakami_PhysRevB.72.085112,
    Koga20051366,%
    Scalettar_PhysRevLett.102.226402,%
    Gebhard_0953-8984-19-43-436206,%
    Peters_PhysRevB.81.035112,%
    Phillips_PhysRevB.84.115101,%
    medici2011janus, *Georges_ANDP:ANDP201100042}.
  We assume semi-elliptic
  densities of states, $\rho_{m}(\epsilon)$ $=$
  $(8/\pi)\sqrt{(W_m/2)^2-\epsilon^2}/W_m^2$, with bandwidths $W_1$
  $<$ $W_2$.  Since the hopping amplitudes $t_{ij,m}$ are diagonal in
  the band index, so are the single-particle Green functions and
  self-energies.  Off-diagonal contributions only occur in
  two-particle and higher-order correlation functions, e.g.,  spin and
  charge susceptibilities.

  In DMFT the model \eqref{eq:LattHam} is mapped onto the following
  two-impurity Anderson model
  (TIAM)~\cite{Koga20051366, *Koga_PhysRevB.73.155106}
  \begin{align}
    H_{\text{TIAM}}
    &=
    \sum_{\bm{k}m\sigma}
    \epsilon_{\bm{k}m}c_{\bm{k}m\sigma}^{\dag}c_{\bm{k}m\sigma}^{\phdag}
    +
    \sum_{m\sigma}
    \epsilon_{m}n_{m\sigma}
    \nonumber\\&~~~
    +
    \sum_{\bm{k}m\sigma}
    \big(
    V_{\bm{k}m}c_{\bm{k}m\sigma}^{\dag}d_{m\sigma}^{\phdag}+\text{h.c.}
    \big)
    +
    H_{\text{int}}^{\text{loc}}
    \,,\label{eq:TIAM}
  \end{align}
  where the local interaction $H_{\text{int}}^{\text{loc}}$ has the
  same form as $H_{\text{int}}$, but without the index $i$.  The DMFT
  self-consistency conditions demand that the band energies
  $\epsilon_{\bm{k}m}$ and hybridizations $V_{\bm{k}m}$ are determined
  such that Green functions and self-energies of~\eqref{eq:TIAM} equal
  the corresponding local lattice quantities,
  \begin{subequations}%
    \begin{align}%
      G_{m}(\omega)
      &=
      \int d\epsilon
      \frac{\rho_{m}(\epsilon)}{\omega+i0-\epsilon_{m}-\Sigma_{m}(\omega)-\epsilon}
      \,,\label{eq:SelfCons}\\
      \Sigma_{m\sigma}(\omega)
      &=
      \omega+i0-\epsilon_{m}-\frac{1}{G_{m}(\omega)}-\Delta_{m}(\omega)
      \,,\label{eq:Delta}
    \end{align}%
    \label{eq:DMFTsc}%
  \end{subequations}%
  which take the same form as for two decoupled one-band models due to
  the absence of interorbital hopping.  Here the hybridization function
  is defined as $\Delta_{m}(\omega)$ $=$
  $\sum_{\bm{k}}|V_{\bm{k}m}|^2/(\omega+i0-\epsilon_{\bm{k}m})$.
  We solve the impurity model using the Numerical Renormalization
  Group (NRG). The complete Fock space NRG~\cite{PhysRevB.74.245114}
  proceeds as in the single-band Hubbard
  model~\cite{PhysRevLett.83.136}, but with a local dimension of 16
  for the impurity and the chain
  sites~\cite{springerlink:10.1140/epjb/e2005-00117-4,PhysRevLett.83.136}.
  Unless noted otherwise the NRG discretization parameter is
  $\Lambda=2.5$, and we keep on the order of $10^{5}$ states,
  including multiplicities of irreducible subspaces, in each NRG
  iteration.  Although we focus on the low-energy region we employ
  Oliveira's~\cite{PhysRevB.49.11986} $z$-trick ($N_{z}=4$) to improve
  the spectra at higher energies. To obtain the self-energy with high
  quality the correlation function $F_{m}(\omega)$ $=$
  $\langle\!\langle[d_{m,\sigma}^{\phdag},H_{\text{int}}^{\text{loc}}];
  d_{m,\sigma}^{\dag}\rangle\!\rangle_{\omega}$ is calculated since it is numerically
  better conditioned than the Dyson
  equation~\cite{bulla1998numerical}.

\begin{figure}[t]
    \centerline{\includegraphics[width=0.9\hsize]{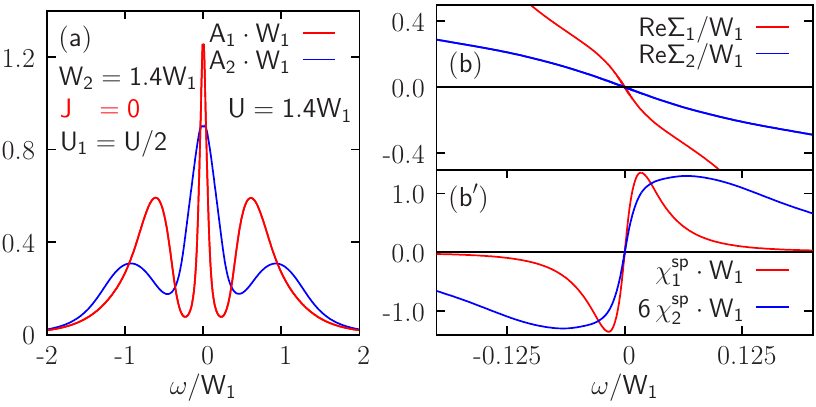}}
    \caption{\label{fig:onlyU1}For $J=0$ and $U_{1}\neq0$ the
      difference in the correlation strength of the two bands is
      observed not only in the shape of the spectral function
      $A_{m}(\omega)$ (a), but also in the corresponding band-resolved
      Kondo temperatures and widths of the Fermi-liquid regime (b,
      b$^{\prime}$).}
  \end{figure}

\begin{figure}[t]
    \centerline{\includegraphics[width=0.9\hsize]{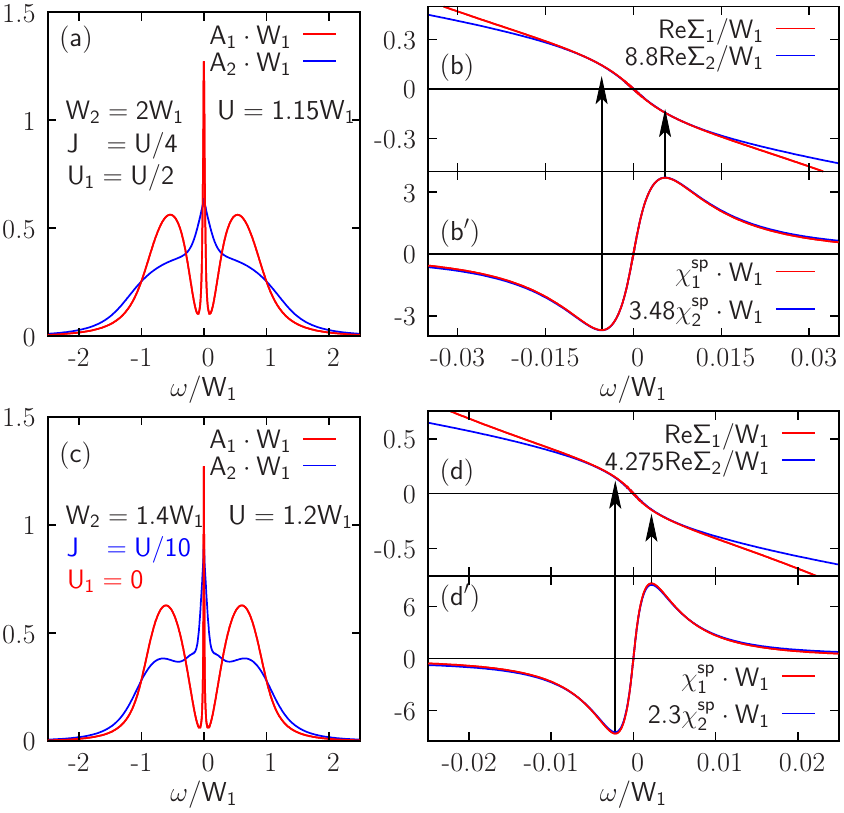}}
    \caption{\label{fig:coupleLowEn}In the metallic phase close to the
      OSMT a striking proportionality, $\text{Re}[\Sigma_{1}]$
      $\propto$ $\text{Re}[\Sigma_{2}]$ and $\chi^{\text{sp}}_{1}$
      $\propto$ $\chi^{\text{sp}}_{2}$ (panels b, b$^{\prime}$, d,
      d$^{\prime}$) is seen to emerge for $J>0$, which is in marked
      contrast to the decoupled behavior in Fig.~\ref{fig:onlyU1}.
      Very close to the OSMT (d, d$^{\prime}$) even a weak exchange
      coupling $J$ leads to a common low-energy scale. Note that $U_1$
      has little influence on the qualitative low-energy behavior.}
  \end{figure}

\begin{figure}[t]
    \centerline{\includegraphics[width=0.82\hsize]{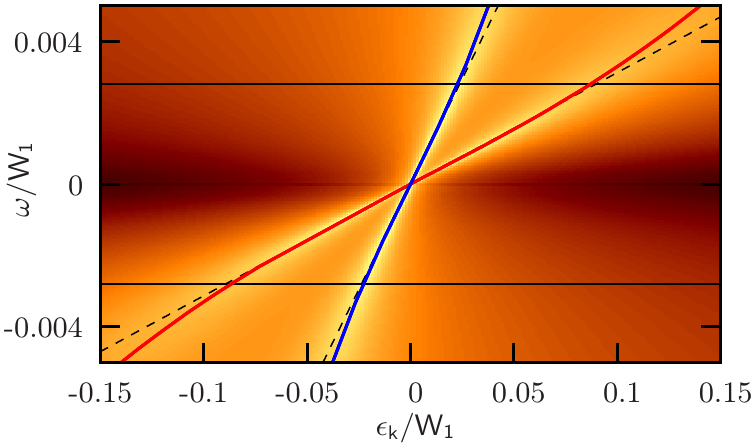}}
    \caption{\label{fig:bandstructure}Intensity plot of the total
      momentum-resolved spectral function $A_{\bm{k}}(\omega)$ deep
      inside the central peaks as a function of $\epsilon_{\bm{k}}$
      and $\omega$ for the same parameters as in
      Fig.~\ref{fig:coupleLowEn}a,b. The effective dispersion
      $E_{\bm{k},m}$ is given by the local maxima in
      $A_{\bm{k}}(\omega)$ (blue line for the wide band, lighter red
      line for the narrow band). It is linear (dashed lines) in the
      Fermi-liquid regime close to the Fermi surface and has kinks at
      the same energy $\sim\pm0.0028W_1$ for both bands (solid
      horizontal lines).}
  \end{figure}

  We compute the spectra, self-energies, and spin susceptibilities for
  the two bands in the metallic phase close to the OSMT, and monitor
  the behavior as a function of the Hund's rule coupling $J$.  The
  different correlation strengths of the orbitals lead to different
  behavior of the spectral functions $A_{m}(\omega)$ $=$
  $-\text{Im}[G_m(\omega)]/\pi$ and self-energies $\Sigma_{m}(\omega)$
  which are shown in Figs.~\ref{fig:onlyU1}
  and~\ref{fig:coupleLowEn}.
  The overall behavior corresponds to that of two Fermi liquids with
  different mass renormalizations, reminiscent of two uncoupled
  one-band Hubbard models with different local interactions.  In all
  cases the single-particle spectra $A_{1}(\omega)$ and
  $A_{2}(\omega)$ differ significantly, especially close to the OSMT
  when the spectral function of the narrow band has a very sharp
  central peak.  Because the spectrum of each band depends mostly on
  the correlation strength $U/W_m$ but not much on Hund's rule
  exchange $J$, the latter was previously characterized as a `band
  decoupler'~\cite{Medici_PhysRevB.83.205112}, at least for weak
  interorbital hopping.  Quantum-Monte Carlo results
  \cite{Koga20051366,Arita_PhysRevB.72.201102,Bluemer_PhysRevB.72.081103,Scalettar_PhysRevLett.102.226402,Medici_PhysRevB.72.205124}
  suggest that at $T=0$ the low-frequency behavior of
  $\text{Re}[\Sigma_{m}]$ and $\chi^{\text{sp}}_{m}$ is then also
  different.  Fig.~\ref{fig:onlyU1} shows that this is indeed the case
  --- but only for $J=0$ and $U_{1}\neq0$.  Indeed, a finite Hund's
  rule coupling leads to a fundamentally different low-energy behavior
  of $\text{Re}[\Sigma_{m}]$ and $\chi^{\text{sp}}_{m}$. Namely, as
  the system approaches the OSMT we find that at low energies these
  quantities become proportional, i.e.,
  $\text{Re}[\Sigma_{1}(\omega)]\propto \text{Re}[\Sigma_{2}(\omega)]$
  and
  $\chi^{\text{sp}}_{1}(\omega)\propto\chi^{\text{sp}}_{2}(\omega)$~\footnote{A
    similar proportionality is found for the imaginary part of the
    self-energy, $\text{Im}[\Sigma_{1}(\omega)]\propto
    \text{Im}[\Sigma_{2}(\omega)]$, in the same interval as for its
    real part. }.  As illustrated in Figs.~\ref{fig:coupleLowEn}b,
  \ref{fig:coupleLowEn}b$^{\prime}$, \ref{fig:coupleLowEn}d,
  \ref{fig:coupleLowEn}d$^{\prime}$ this striking result cannot be
  inferred from the spectral functions $A_{m}(\omega)$ since the shape
  and the characteristic energy scales of the latter differ
  considerably and thus suggest a decoupled behavior.  The
  characteristic energy scale of the spin fluctuations, i.e., the
  locations $\omega^{\text{sp}}_{m}$ of the extrema in the spin
  susceptibilities $\chi^{\text{sp}}_{m}(\omega)$, allow one to define
  a Kondo temperature for each band.  The proportionalities discussed
  above imply that the system has identical Kondo temperatures
  ($\omega^{\text{sp}}_{1}=\omega^{\text{sp}}_{2}$), Fermi-liquid
  energy scales and kinks ($\omega^{\Sigma}_{1}=\omega^{\Sigma}_{2}$),
  \emph{irrespective} of the different correlation strengths of the
  bands.  Furthermore, the self-energy kinks and the strongest spin
  fluctuations occur at the same energy in each band,
  $\omega^{\Sigma}_{m}\simeq\omega^{\text{sp}}_{m}$ (as observed also
  in the single-band case~\cite{PhysRevLett.102.076406}), which means
  that for the two-band model~\eqref{eq:LattHam} a \emph{single common
    low-energy scale} emerges for kinks and spin fluctuations in both
  bands.  For our numerical data we define the kink scale
  $\omega^{\Sigma}_{m}$ as the energy for which the extrapolated
  linear dispersion near the Fermi energy deviates from
  $\text{Re}[\Sigma_{m}(\omega)]$ by 20\%, which agrees well with the
  perceived location of the kinks in Figs.~\ref{fig:coupleLowEn}b,d.
  The corresponding momentum-resolved spectral function
  $A_{\bm{k}}(\omega)$ and effective dispersions $E_{\bm{k}m}$ are
  shown Fig.~\ref{fig:bandstructure}.
  We observe that although the slope of the Fermi-liquid dispersion is
  very different for the two bands, the linear regimes terminate at
  the \emph{same} energy scale, which however slightly deviates
  from $\omega^{\Sigma}_{m}$ due to band structure effects.

\begin{figure}
    \centerline{\includegraphics[width=0.8\hsize]{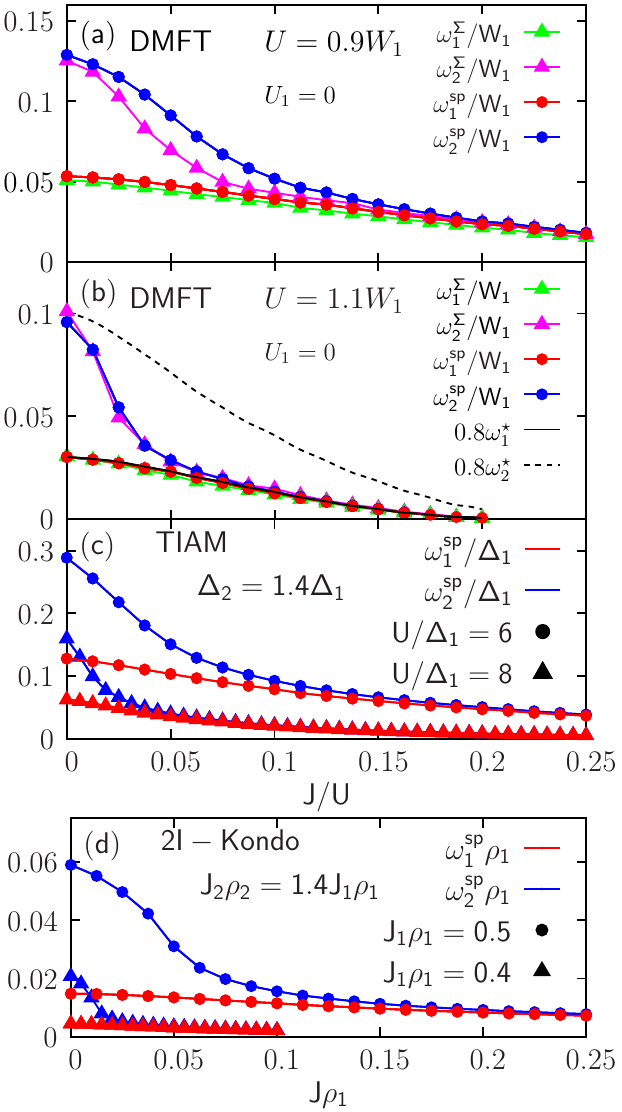}}
    \caption{\label{fig:KinkschiofJ} (a,b) Orbitally resolved kink
      energies $\omega_{1,2}^{\Sigma}(J)$ and Kondo temperatures
      $\omega_{1,2}^{\text{sp}}(J)$ as calculated within DMFT for
      different Hubbard interactions $U$.  Sufficiently large $U$
      leads to $\omega_{m}^{\Sigma}(J)=\omega_{m}^{\text{sp}}(J)$ and
      sufficiently large $J$ leads to
      $\omega_{1}^{\Sigma}(J)=\omega_{2}^{\Sigma}(J)$ and
      $\omega_{1}^{\text{sp}}(J)=\omega_{2}^{\text{sp}}(J)$. The
      dashed lines in (b) mark the single-band estimate
      for $\omega_{m}^{\Sigma}$~\cite{byczuk2006kinks}
      which applies only to
      the narrow band (see text).
      (c) A simplified Anderson impurity model and, (d), a related
      two-impurity Kondo model, both with a behavior similar to the
      DMFT solution.  Note that the system in (b) enters the OSMT
      phase at approximately $J=0.2U$. For these very low-energy
      features $\mathcal{O}(5500)$ states were kept and $N_{z}=2$.}
  \end{figure}

  By comparing the results for $J=0$ in Fig.~\ref{fig:onlyU1} with
  $J\neq0$ in Figs.~\ref{fig:coupleLowEn}b-c it is clear that the
  interorbital repulsion $U_{1}$ is not responsible for the common
  energy scale.  This effect only appears in the presence of the
  Hund's rule coupling $J$, whereas $U_{1}$ merely leads to
  quantitative modifications.  We will thus restrict ourselves to
  $U_{1}=0$ in the following.
  Starting from $J=0$ we study the continuous evolution of the two
  initially uncoupled Hubbard models into the `locked' regime. To this
  end we obtain Kondo scales $\omega^{\text{sp}}_{m}(J)$ and
  self-energy kinks $\omega^{\Sigma}_{m}(J)$ for the two orbitals for
  two different values of $U/W_1$ (Figs.~\ref{fig:KinkschiofJ}a-b).
  As $J$ is increased both orbital-resolved energy scales approach
  each other and finally merge into a single scale, as seen in
  Figs.~\ref{fig:coupleLowEn}c-d$^{\prime}$.  Comparing
  Figs.~\ref{fig:KinkschiofJ}a and~b, we observe that this common
  low-energy scale appears at a threshold value which decreases for
  increasing $U$.  We also notice the very close correspondence
  between kink energies and Kondo-temperatures, especially in the more
  strongly correlated case (Fig.~\ref{fig:KinkschiofJ}b).  This
  observation can be understood in terms of local Fermi-liquid theory
  \cite{springerlink:10.1007/BF00654541}: Since the binding energy of
  the Kondo singlet is approximately given by the Kondo temperature,
  the linear regime must terminate at
  $\omega\simeq\omega_{\text{sp}}$; see
  Refs.~\cite{PhysRevLett.102.076406,grete2011kinks} for a discussion
  of the single-band case.  As expected the kink energy scale derived
  in Ref.~\cite{byczuk2006kinks}, $\omega^{\star}_{m}$ $=$
  $0.2Z_mW_m$, applies only to the narrow band with its well-developed
  three-peak spectral function and not to the wide band (cf.\ solid
  and dashed lines, respectively, in Fig.~\ref{fig:KinkschiofJ}b,
  corresponding to $0.8\omega^{\star}_{m}$~\footnote{The prefactor of
    0.8 accounts for our present kink criterion (20\% deviation
    in the self-energy from the linear extrapolation, see text), which
    differs slightly from that in Ref.~\cite{byczuk2006kinks}.}).

  In the vicinity of the OSMT the characteristic energies
  $\omega^{\text{sp}}_{m}(J)$ and $\omega^{\Sigma}_{m}(J)$ represent
  equivalent energy scales and hence contain the same physical
  information.  In order to explain the $J$ dependence it suffices to
  discuss one of them, and we will focus on
  $\omega^{\text{sp}}_{m}(J)$ in the following. To explain the locking
  of the low-energy scales for the two bands we proceed in two steps;
  see Figs.~\ref{fig:KinkschiofJ}c and \ref{fig:KinkschiofJ}d. First
  we establish that the locking is an intrinsic property of the
  underlying TIAM Hamiltonian and is only quantitatively modified by
  the DMFT self-consistency equations~\eqref{eq:DMFTsc}. Then we
  compare with the results for a Kondo-type model that allows us to
  identify the competing couplings and elementary excitations.  For
  the first step we solve the impurity model~\eqref{eq:TIAM} with
  different but \emph{constant} hybridization functions
  ($\Delta_{2}(\omega)$ $=$ $1.4\Delta_{1}(\omega)$ $=$ const) and
  extract the Kondo temperatures $\omega^{\text{sp}}_{m}(J)$ from the
  maxima of the spin susceptibilities.  The result is depicted in
  Fig.~\ref{fig:KinkschiofJ}c for two values of $U$, showing very good
  qualitative agreement with the DMFT results in
  Figs.~\ref{fig:KinkschiofJ}a, \ref{fig:KinkschiofJ}b. In particular,
  the common low-energy scale emerges at a value of $J$ which
  decreases with increasing $U$ in a similar fashion.  We conclude
  that the DMFT self-consistency induces only minor modifications as
  long as the system remains in the metallic phase.

  In a second step we focus on the low-energy spin dynamics close to
  the OSMT. In this regime charge excitations are strongly suppressed.
  We thus consider the Kondo limit of~\eqref{eq:TIAM} which captures
  the low-energy spin dynamics of the TIAM in
  Fig.~\ref{fig:KinkschiofJ}c. In this limit the
  Hamiltonian~\eqref{eq:TIAM} reduces to a two-impurity Kondo model
  (2IKM)~\cite{Jayaprakash1981TwoImpurity,Jones1987Study},
  \begin{multline}
    H_{\text{2IKM}}
    =
    \sum_{\bm{k}m\sigma}
    \epsilon_{\bm{k}m}
    c_{\bm{k}m\sigma}^{\dag}
    c_{\bm{k}m\sigma}^{\phdag}
    \\
    +
    \sum_{m}J_{m}
    \bm{s}_{m}\cdot\bm{S}_{m}
    -
    J\bm{S}_{1}\cdot\bm{S}_{2}
    \,.\label{eq:Kondo}
  \end{multline}
  Here $J$ $>$ $0$ is the Hund's exchange interaction
  of~\eqref{eq:TIAM}, while the antiferromagnetic couplings $J_m$ stem
  from superexchange processes and decrease with increasing $U$.  We
  take $J_{2}\rho_{2}(0)=1.4\, J_{1}\rho_{1}(0)$ to obtain different
  Kondo temperatures for $J=0$, i.e.,
  $\omega^{\text{sp}}_{1}(0)\neq\omega^{\text{sp}}_{2}(0)$. The $J$
  dependence of $\omega^{\text{sp}}_{m}(J)$ is shown in
  Fig.~\ref{fig:KinkschiofJ}d. The qualitative agreement among the
  results obtained for all three models (DMFT, TIAM with constant
  hybridization, 2IKM) confirms that~\eqref{eq:Kondo} already
  describes the essential processes that lead to the emergence of the
  joint low-energy scale. In the 2IKM the spins will align for
  low-excitation energies and form a composite spin-1
  object~\cite{Jayaprakash1981TwoImpurity}. This happens roughly when
  the energy gain
  $\omega^{\text{sp}}_{1}(J)+\omega^{\text{sp}}_{2}(J)$ due to Kondo
  screening of the two impurities is overcome by $J/4$, the
  approximate energy gain due to the ferromagnetic exchange. Hence the
  locking of the low-energy scales sets in at about $J$ $\approx$
  $\omega^{\text{sp}}_{m}(J)/8$, as seen in
  Fig.~\ref{fig:KinkschiofJ}a ($J\approx0.2W_1$)
  and~\ref{fig:KinkschiofJ}b ($J\approx0.1W_1$).  We conclude that the
  low-energy spin dynamics of the two impurity spins (and thus the two
  bands of the corresponding lattice model in DMFT) exhibit joint
  fluctuations and thus have equal Kondo scales if $J$ 
  dominates the individual Kondo scales $\omega^{\text{sp}}_{m}$
  and are essentially independent otherwise.  Regarding the influence
  of $U$, we note that the antiferromagnetic couplings between the
  spins and the baths decrease with increasing interaction, i.e., more
  correlated systems exhibit stronger locking of their spins and their
  low-energy scales.  Thus $J$ couples the low-energy scales more
  effectively for stronger correlations, as seen in
  Fig.~\ref{fig:KinkschiofJ}a-c.

\begin{figure}[t]
    \centerline{\includegraphics[width=0.9\hsize]{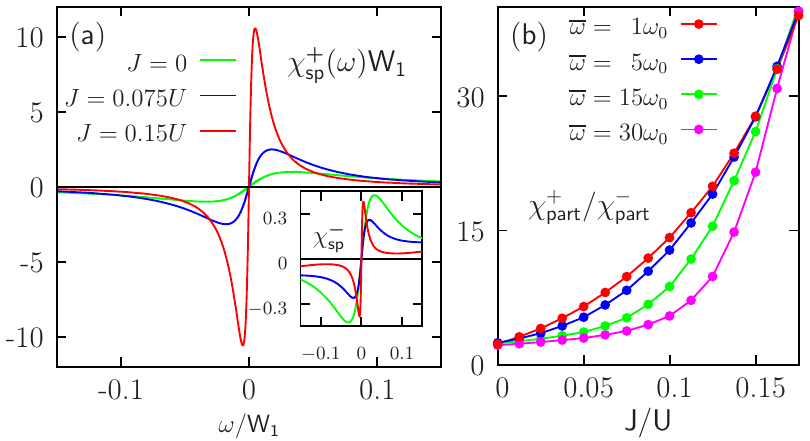}}
    \caption{\label{fig:SPlusSMinusCorrFun}The correlation functions
      $\chi_{\text{sp}}^{+}(\omega)$ and
      $\chi_{\text{sp}}^{-}(\omega)$ indicate the creation of a
      composite spin-1 object for large $J$ (green curves in (a) and
      in the inset have about the same magnitude; also note the
      different scales in (a) and the inset). Accordingly, the partial
      spectral weight fraction $\chi_{part.}^{+}/\chi_{part.}^{-}$ in
      (b) grows strongly with $J$ and is only quantitatively affected
      by the upper limit $\overline{\omega}$. Here $\omega_{0}$ $=$
      $\max[\omega^{\text{sp}}_{1}(J),\omega^{\text{sp}}_{2}(J)]$.}
  \end{figure}

  To explicitly verify the physical picture described above we
  investigate the correlation functions
  \begin{align}
    \chi_{\text{\text{sp}}}^{\pm}(\omega)
    &=
    \langle\!\langle\bm{S}_{1}\pm\bm{S}_{2}
    |\bm{S}_{1}\pm\bm{S}_{2}\rangle\!\rangle_{\omega}
    \label{eq:spin_one_plus_minus}
  \end{align}
  in DMFT.  They describe the dynamics of the composite spin-1 object
  $\chi^{+}(\omega)$ and the `residual' singlet $\chi^{-}(\omega)$,
  respectively, and are plotted in Fig.~\ref{fig:SPlusSMinusCorrFun}a.
  As expected, for both susceptibilities the positions of their maxima
  decreases with increasing interaction. However, only the composite
  spin-1 object shows a resonance that also increases in amplitude.
  In Fig.~\ref{fig:SPlusSMinusCorrFun}b we plot the integrated weight
  $\chi_{part}^{\pm}$ $=$
  $\int_{0}^{\overline{\omega}}\chi^{\pm}(\omega)d\omega$ as a
  function $J$ for several values of the upper limit
  $\overline{\omega}$.  In particular, for larger values of $J$ the
  residual spin contributes only little to the total low-energy spin
  response $\chi^{\text{sp}}_{1}+\chi^{\text{sp}}_{2}$ of the system,
  which is thus well described by $\chi^{+}(\omega)$. This establishes
  the formation of the composite spin-1 object as the physical origin
  for the emergence of the common energy scale.  Namely, as the OSMT
  is approached by increasing $U$, the antiferromagnetic superexchange
  of the narrow and wide band both decrease until $J$ becomes the
  dominating scale, at least for the spins in the narrow band. The
  spins align and create a composite spin-1 and exhibit joint
  low-energy dynamics, leading to the proportionalities of
  $\text{Re}[\Sigma_{m}(\omega)]$ and $\chi^{\text{sp}}_{m}(\omega)$
  for the two bands.

  In summary, we explored the physical mechanism for kinks responsible
  for the appearance of kinks in the self-energy by studying a
  two-band model with different bandwidths but the same local charge
  interactions as well as Hund's rule spin exchange.  We find that the
  physical picture developed previously for single-band systems close
  to the Mott transition is significantly modified for strong Hund's
  rule coupling, due to the formation of a local spin-1 object.  As a
  consequence, a common low-energy scale emerges for the two bands for
  the kinks in the self-energies and the maxima in the spin
  susceptibilities.

  We thank R.~\v{Z}itko for releasing his NRG code, on which our code
  is based \cite{Rok20112259}, and L.~Chioncel and J.~Otsuki
  for useful discussions. This work was supported by the Deutsche
  Forschungsgemeinschaft through TRR~80.

\end{document}